\documentclass[a4paper, nofootinbib]{revtex4}
\pdfoutput=1
\usepackage{graphicx}
\usepackage{fancyhdr}
\usepackage{amsmath}
\usepackage{color}

\begin{document}

\title{Quantum corrections to the spin-independent cross section in the inert
Higgs doublet model}
\thanks{This talk is based on Ref.~\cite{1501.04161}.}
%

\author{Tomohiro Abe}
\thanks{speaker}
\author{Ryosuke Sato}
\affiliation{
Institute of Particle and Nuclear Studies,
High Energy Accelerator Research Organization (KEK), Tsukuba 305-0801, Japan}

\begin{abstract}
The inert Higgs doublet model is an extension of the standard model with
 an extra 
scalar doublet. The extra doublet is $Z_2$ odd while all the other
particles are $Z_2$ even, thus the model contains a dark matter
candidate. It is known that the spin-independent cross section of the
dark matter and the nucleon is highly suppressed for $m_{\text DM} \sim
 m_h/2$ at the tree level. In this talk, we show the loop corrections give the
significant effects in this regime.
\end{abstract}

\maketitle

\thispagestyle{fancy}


\section{Introduction}

One of the reasons why we believe in a model beyond the standard
model (SM) is the existence of the dark matter (DM) in our universe.
The inert doublet model~\cite{PHRVA.D18.2574, hep-ph/0603188} is a relatively simple model 
among the models with the DM. 
In this model, it is known that there are two viable dark
matter mass regions, 53~GeV $\lesssim m_{\text{DM}} \lesssim 71$~GeV and
$m_{\text{DM}} \gtrsim$ 500~GeV~\cite{LopezHonorez:2006gr, JCAP.1406.030, Abe:2014gua} under the assumption of the thermal
relic abundance scenario \cite{PRLTA.39.165, NUPHA.B310.693, NUPHA.B360.145}.

In the lighter dark matter case, the dark matter almost annihilate
into $b\bar{b}$ as shown in the left panel in
 Fig.~\ref{fig:annihilation-diagrams}. 
When the dark matter mass is less than half of the Higgs boson mass, the
 annihilation cross section of this process is 
 enhanced due to the Higgs resonance, and the amount of the dark matter
 relic abundance getting smaller than the current observed amount of the
 dark matter \cite{Ade:2013zuv} unless 
 the DM-Higgs coupling becomes small. Therefore, $\lambda_A$ should be highly
 suppressed in the light dark matter mass regime to explain the observed
 value of the dark matter energy density. 
 The important consequence of the highly suppressed $\lambda_A$ is the
 very small spin-independent cross section of the dark matter and
 the nucleon. The spin-independent cross section is smaller than $10^{-47}$~cm$^2$ for
 57~GeV $\lesssim m_{\text DM} \lesssim 63$~GeV.
 This makes direct search for the dark matter difficult.

It was pointed out that the quantum corrections via the gauge bosons are
significant and the spin-independent cross section is enhanced when the
cross section at the tree level is very small \cite{PHRVA.D87.075025}.
However, the effects from the self-coupling of the inert-doublet
and the
sign ambiguity of the DM-Higgs coupling were not discussed. 

In this talk, we revisit the quantum corrections to the spin-independent
cross section in the light dark matter mass regime. We take into account
all the relevant diagrams and show the
spin-independent cross section highly depends on the self-coupling of
the inert-doublet and the sign of the DM-Higgs coupling.

\section{Model}

The inert doublet model\footnote{A dynamical realization of this model
was recently discussed in Ref.~\cite{Carmona:2015haa}.} includes two scalar fields ($H$ and $\Phi$)
which are SU(2) doublets with $Y=1/2$. These fields have $Z_2$ symmetry,
\begin{align}
 H \to  H
, \quad
 \Phi \to - \Phi
.
\end{align}
All the other SM fields are unchanged under the $Z_2$ symmetry.
We assume that $\Phi$ does not get non-zero vacuum expectation value (VEV).
We identify $H$ as the SM Higgs field, and $\Phi$ as the \textit{inert} doublet.
The new particles are two neutral scalar bosons ($S$ and $A$) and one
charged Higgs boson ($H^{\pm}$). 
All these particles are the components of $\Phi$.
The lightest particle among them is stable due to the
$Z_2$ symmetry, thus a neutral scalar is a dark matter candidate
as long as the charged Higgs is not the lightest $Z_2$-odd particle.
We can always interchange $S$ and $A$ by the field redefinition of
$\Phi$. Therefore $A$ is generally the dark matter candidate. 
In the following, we consider the situation that $A$ is the lightest
$Z_2$-odd particle and is the dark matter candidate.

There are five new parameters originated from the Higgs
potential. Three of them are the masses of the new particles, 
$m_{A} (=m_{\text{DM}}), m_{S}, m_{H^{\pm}}$, one is the DM coupling to
the SM Higgs boson, $\lambda_A$, and the other is the self-coupling of
the $Z_2$ odd particles, $\lambda_2$.

\section{loop corrections to the cross sections}
\begin{figure}[tb]
  \includegraphics[width=0.20\hsize]{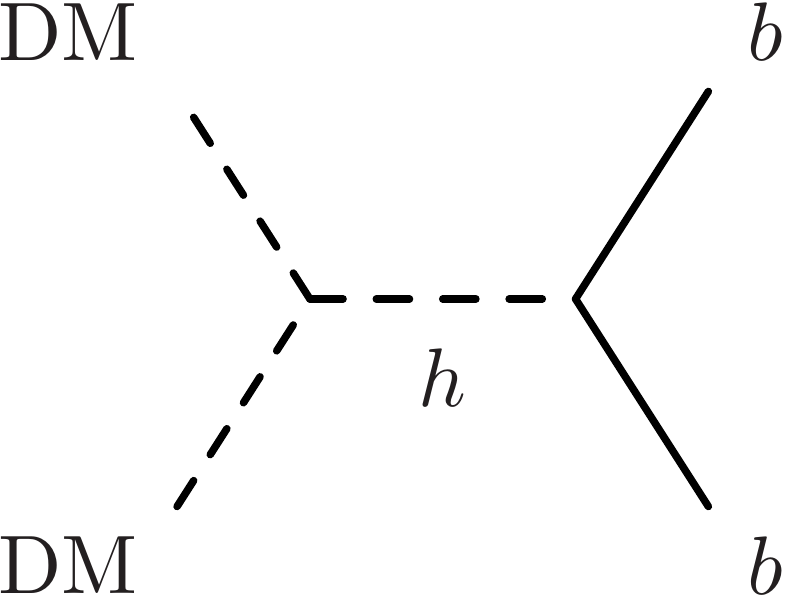}
\quad
  \includegraphics[width=0.20\hsize]{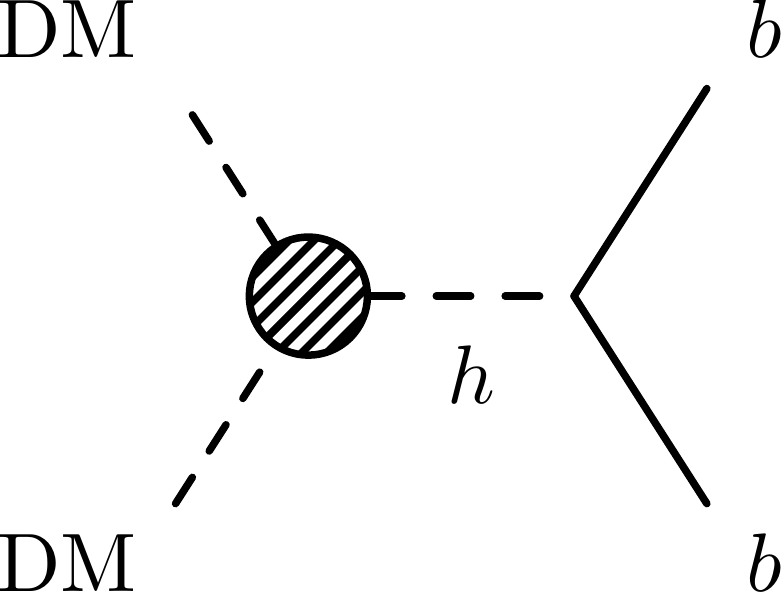}
\caption{ 
Diagrams for the dark matter annihilation cross section. The black
 shaded region means one-loop corrections. 
}
\label{fig:annihilation-diagrams}
\end{figure}

Before discussing the quantum corrections to the spin-independent cross
section, we have to revisit the dark matter annihilation cross
section with loop corrections, because we determine $\lambda_A$ through
the annihilation cross section. We are interested in the dark matter
mass region where the annihilation cross section is dominated by the
Higgs resonance, thus we consider the diagrams shown in
Fig.~\ref{fig:annihilation-diagrams}. The cross section is proportional
to 
\begin{align}
 \sigma v
\propto
\left|
\frac{\lambda_A + \delta \Gamma_h(s) + \delta_{\lambda_A}}{
s - m_h^2 + i m_h \Gamma_h
}
\right|^2
,
\end{align}
where
$\delta \Gamma_h(s)$ is the loop correction shown in the right panel in
Fig.~\ref{fig:annihilation-diagrams}, and $\delta_{\lambda_A}$ is the
counterterm. The loop correction depends on $s$. Since the dominant
contributions to the annihilation cross section come from $s \simeq
m_h^2$, we take on-shell renormalization condition, namely
$\delta_{\lambda_A} = - \delta \Gamma_h(m_h^2)$. Thus almost all the
corrections are absorbed into the counterterm, and we find $\langle \sigma v
\rangle \propto |\lambda_A|^2$. As a result, we can still use the $\lambda_A$
determined at the tree level for our purpose. 

Now $\lambda_A$ is determined from the annihilation cross section. But
what is actually determined is its absolute value. Thus the sign of
$\lambda_A$ is
still unknown, namely $\lambda_A$  has two solutions, $\pm |\lambda_A|$.
This sign ambiguity is not important at the tree level analysis, but 
it is important for the analysis beyond the tree level, because
there are interference terms between the tree diagram and the loop
diagrams. 
We consider both sign possibilities, namely both positive and negative
$\lambda_A$.

\begin{figure}[tb]
  \includegraphics[width=0.13\hsize]{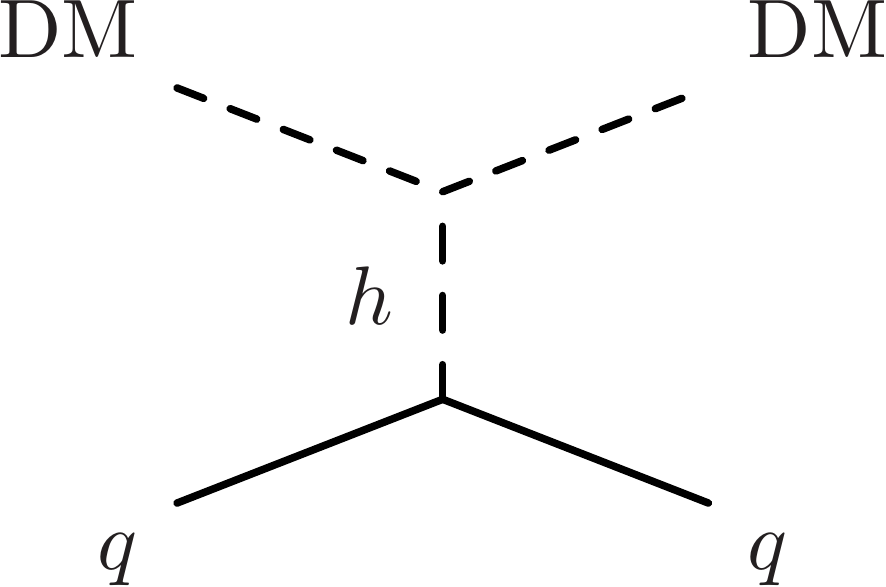}
  \includegraphics[width=0.13\hsize]{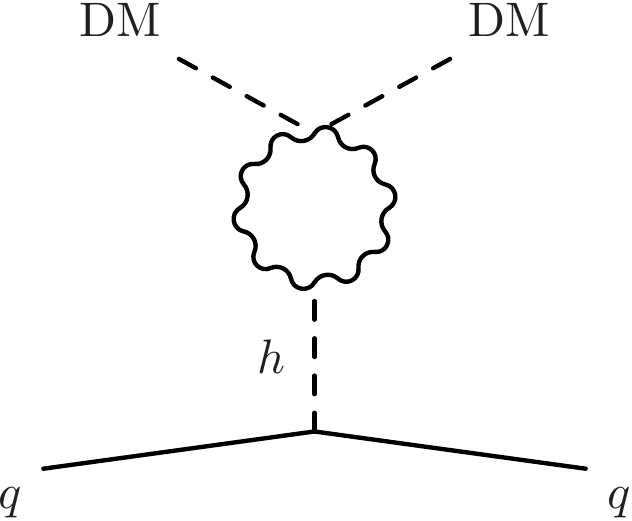}
  \includegraphics[width=0.13\hsize]{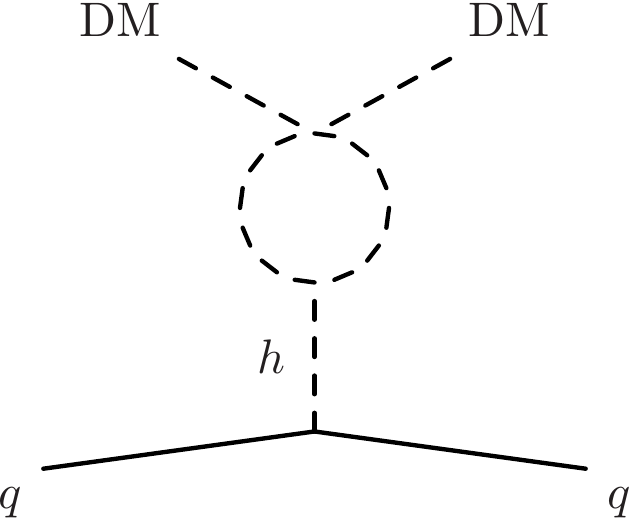}
  \includegraphics[width=0.13\hsize]{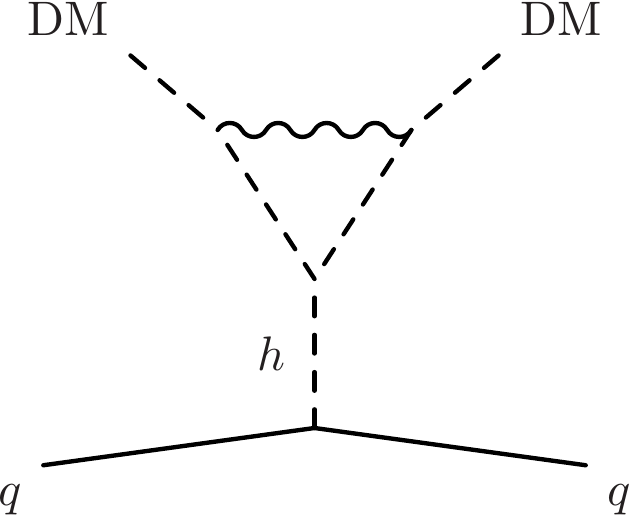}
  \includegraphics[width=0.13\hsize]{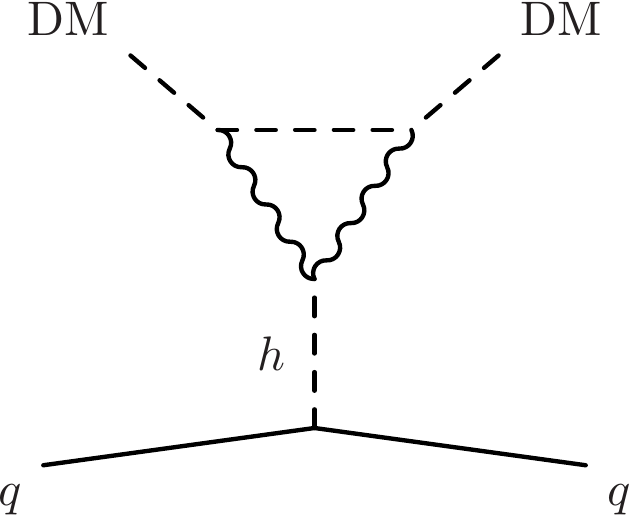}
  \includegraphics[width=0.13\hsize]{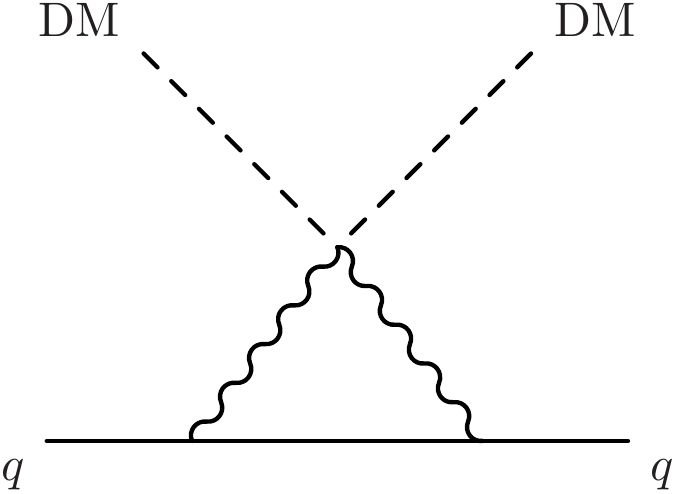}
  \includegraphics[width=0.13\hsize]{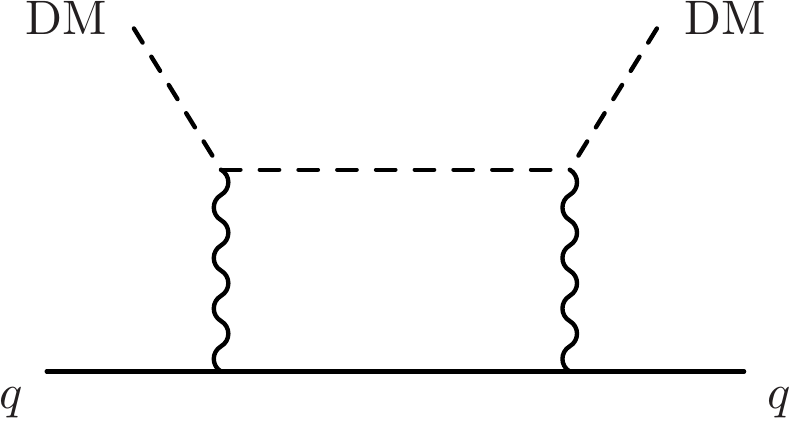}
\\
\vspace{5mm}
  \includegraphics[width=0.13\hsize]{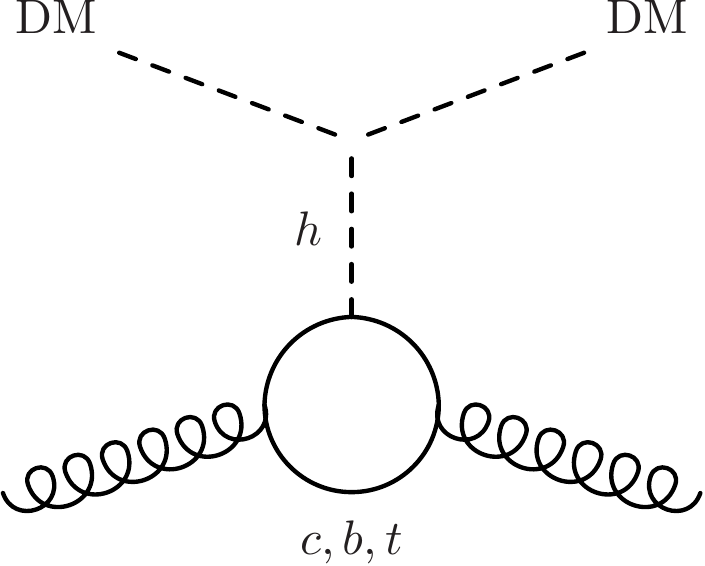}
  \includegraphics[width=0.13\hsize]{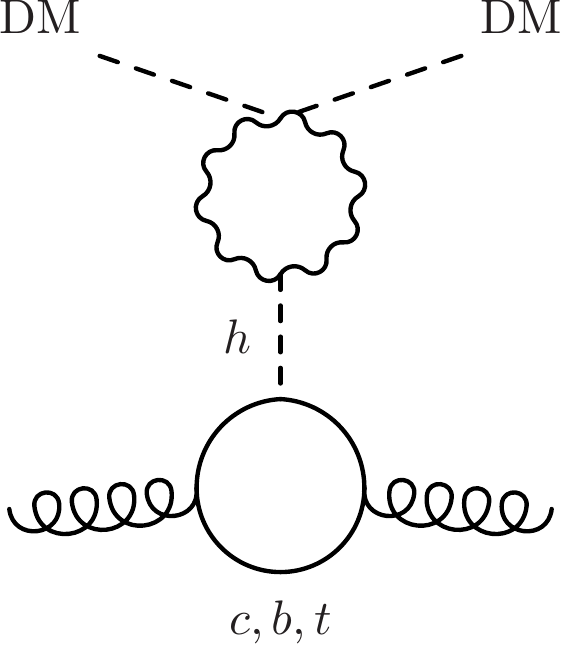}
  \includegraphics[width=0.13\hsize]{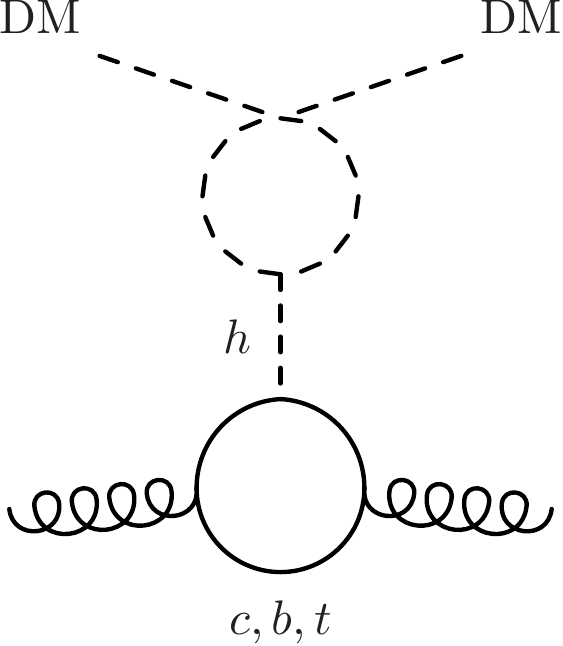}
  \includegraphics[width=0.13\hsize]{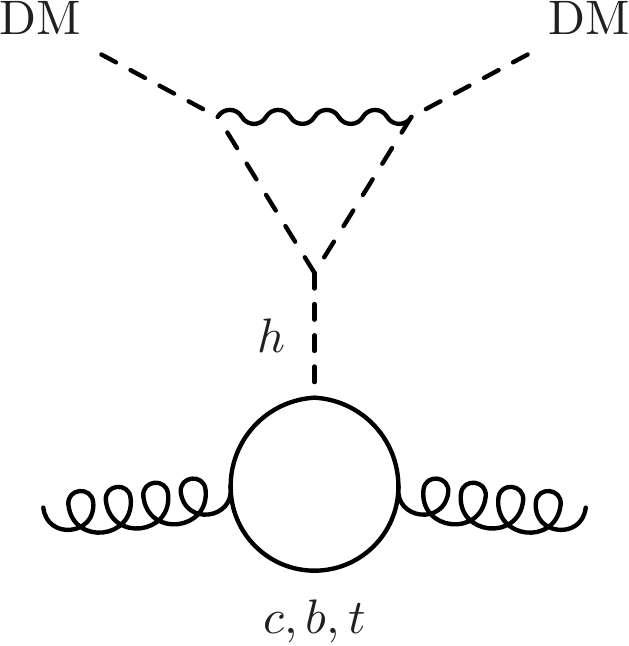}
  \includegraphics[width=0.13\hsize]{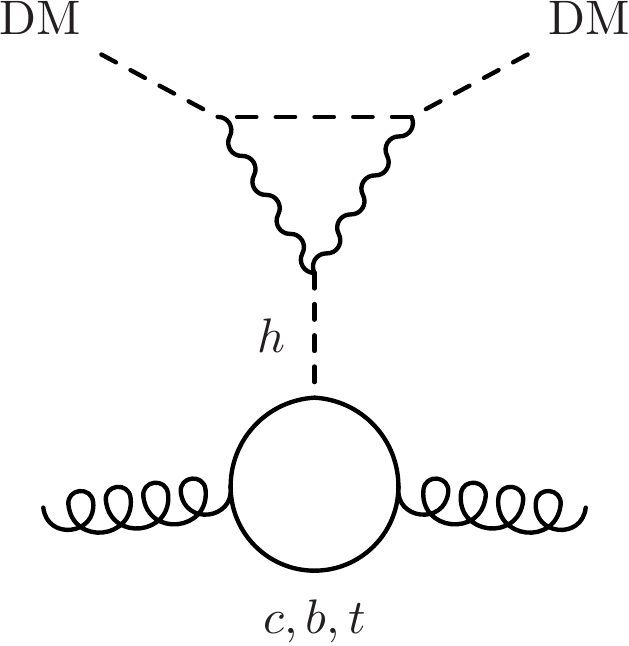}
  \includegraphics[width=0.13\hsize]{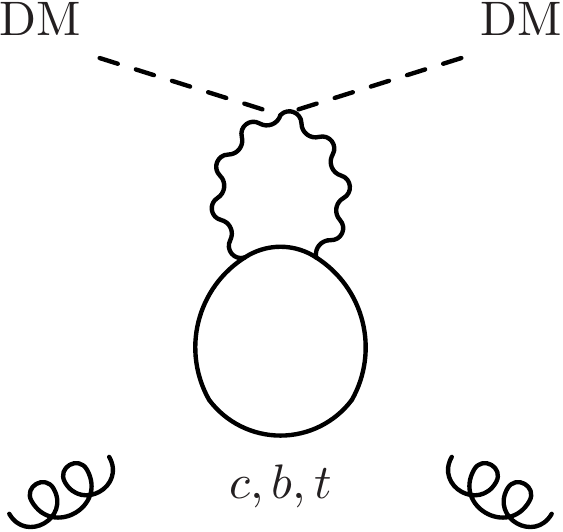}
  \includegraphics[width=0.13\hsize]{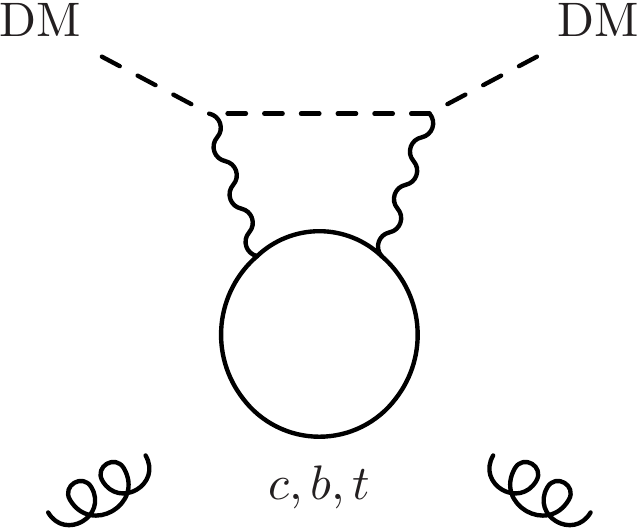}
\caption{
Diagrams for the spin-independent cross section. The DM-quark interaction
 diagrams are in the first line, and the DM-gluon interaction diagrams
 are in the second line. The diagrams in the most left give leading
 order contribution if $\lambda_A$ is not suppressed.
}
\label{fig:xsec-diagrams}
\end{figure}

We move to discuss the spin-independent cross section with the loop
corrections shown in Fig.~\ref{fig:xsec-diagrams}. The spin-independent
cross section is given as
\begin{align}
 \sigma_{\text{SI}}
=
\frac{1}{4\pi}
\frac{
\left( \pm|\lambda_{A}| + \delta \lambda \right)
\mu^2 m_N^2 f_N^2
}{m_A^2 m_h^4}
,
\end{align}
where
$m_N$ is the nucleon mass, $\mu$ is the reduced mass in the dark matter
and nucleon system, $f_N$ is the form factor, and $\delta \lambda$ is
the loop correction given as 
\begin{align}
 \delta \lambda 
=&
 \delta \Gamma_h(0) + \delta_{\lambda_A}
+ 
\text{(box and gluon diagrams)}
.
\end{align}
The counterterm is already determined by the annihilation cross section.
The momentum transfer in the spin-independent cross section is zero, and
different from the annihilation cross section. Thus the total
contribution from the vertex correction is $\delta \Gamma_h(0)
- \delta \Gamma_h(m_h^2)$. For the contributions from the box and the
gluon diagrams, see Ref.\cite{1501.04161}. In the case of $m_{H^{\pm}} = m_S$,
we find the fitting formula for $\delta \lambda$,
\begin{align}
 \delta \lambda 
=&
-0.00409
 \frac{m_{\text{DM}}}{\text{GeV}}
\left( 0.0000144
 - 7.77 \times 10^{-8} \frac{m_{H^{\pm}}}{\text{GeV}} 
 - 0.00334 \frac{\text{GeV}}{m_{H^{\pm}}} 
\right)
\nonumber\\
&+ 
\lambda_2 
\left(
0.00183
-  
7.87 \times 10^{-10} \frac{m_{H^{\pm}}^2 }{\text{GeV}^2}
+ 
\frac{m_{\text{DM}}^2 }{\text{GeV}^2}
\left(-4.13 \times 10^{-8} - 0.00113 \frac{\text{GeV}^2}{m_{H^{\pm}}^2}
\right)
\right)
.
\end{align}

\begin{figure}[tb]
  \includegraphics[width=0.32\hsize]{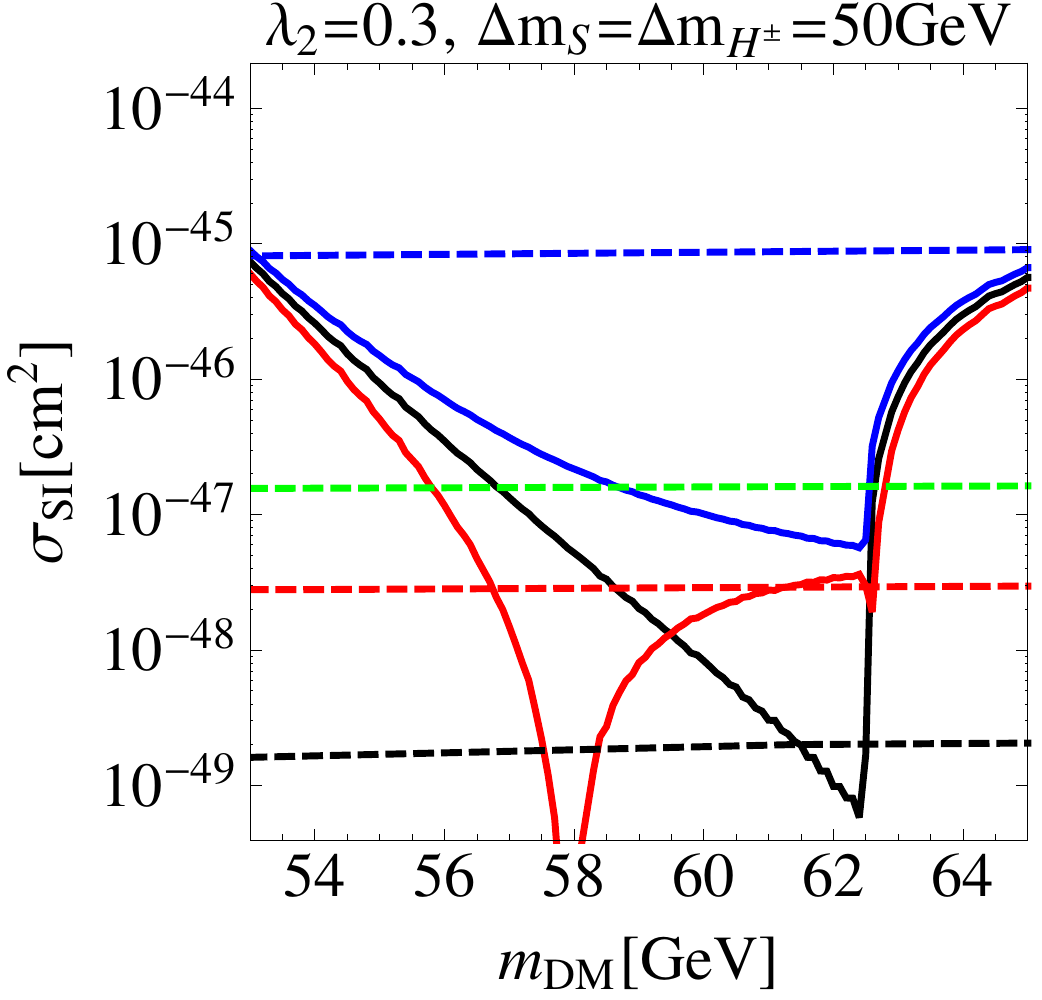} 
  \includegraphics[width=0.32\hsize]{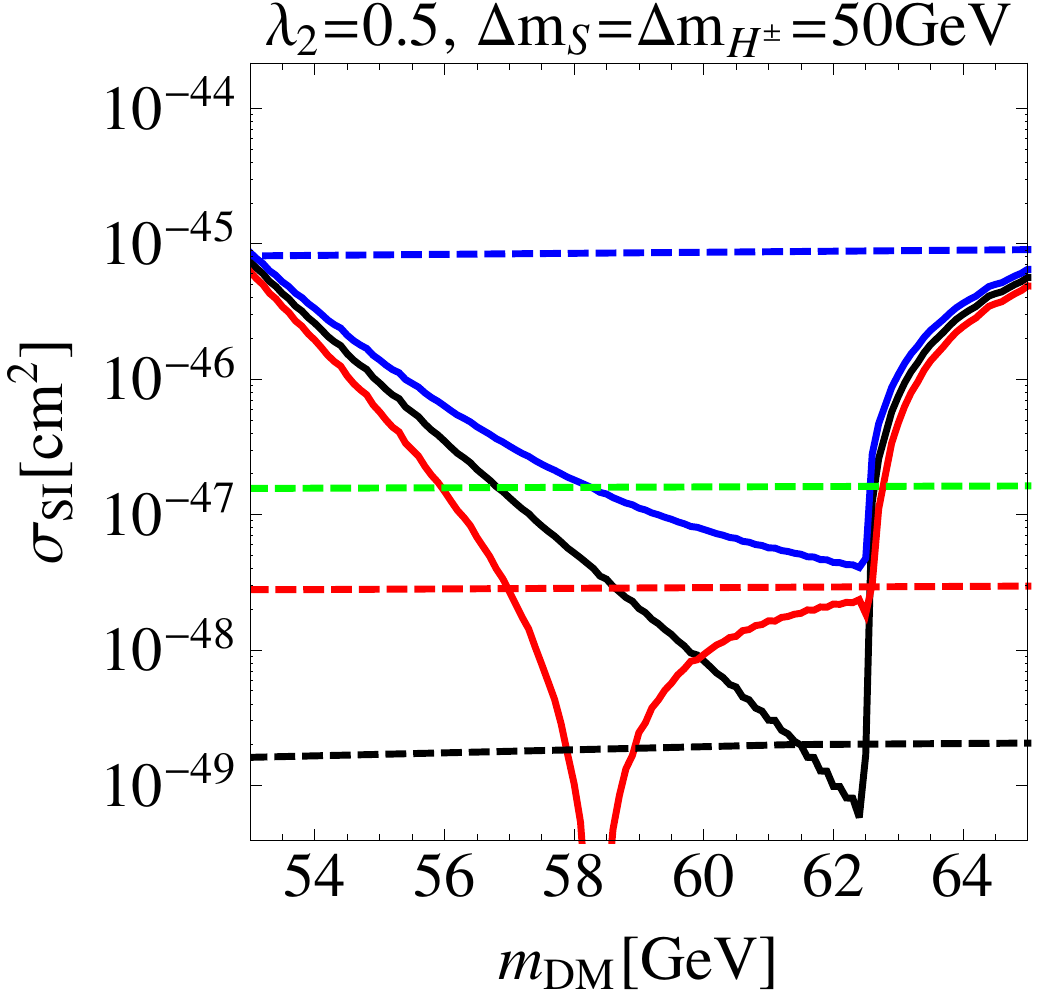} 
  \includegraphics[width=0.32\hsize]{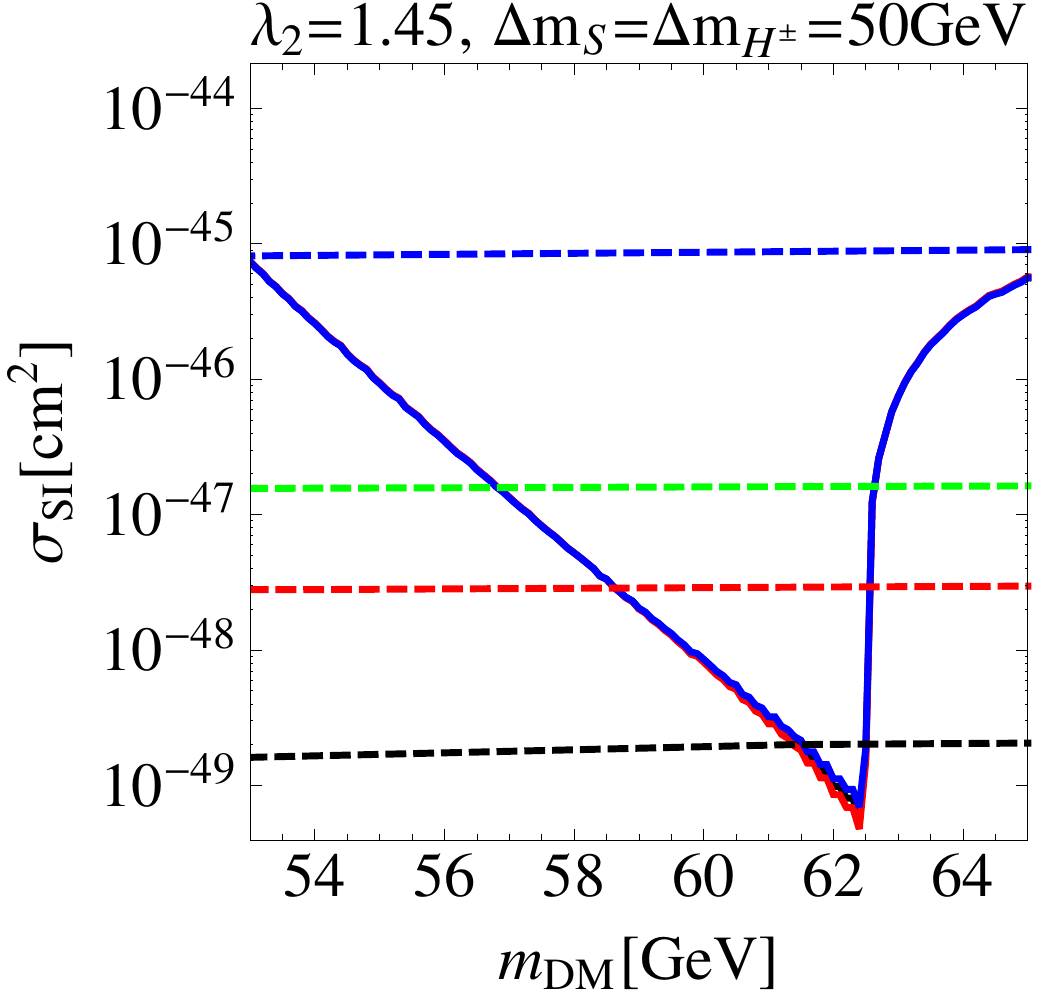} 
\caption{
The spin-independent cross section as a function of the DM mass. The
 black line is the result without the loop corrections. The red (blue)
 line is with the loop corrections and the sign of $\lambda_A$ is
 positive (negative). 
The blue-dashed line is the current LUX bound \cite{1310.8214}. The green-dashed,
 red-dashed lines are the future prospect by XENON1T
 \cite{Aprile:2012zx} and LZ \cite{Feng:2014uja},
 respectively, and the black-dashed line is the discovery limit caused
 by atmospheric and astrophysical neutrinos \cite{Billard:2013qya}.
Here we take $m_{H^{\pm}} = m_{S} = m_{\text{DM}}
 +$ 50~GeV.
}
\label{fig:result1}
\end{figure}

\begin{figure}[tb]
  \includegraphics[width=0.32\hsize]{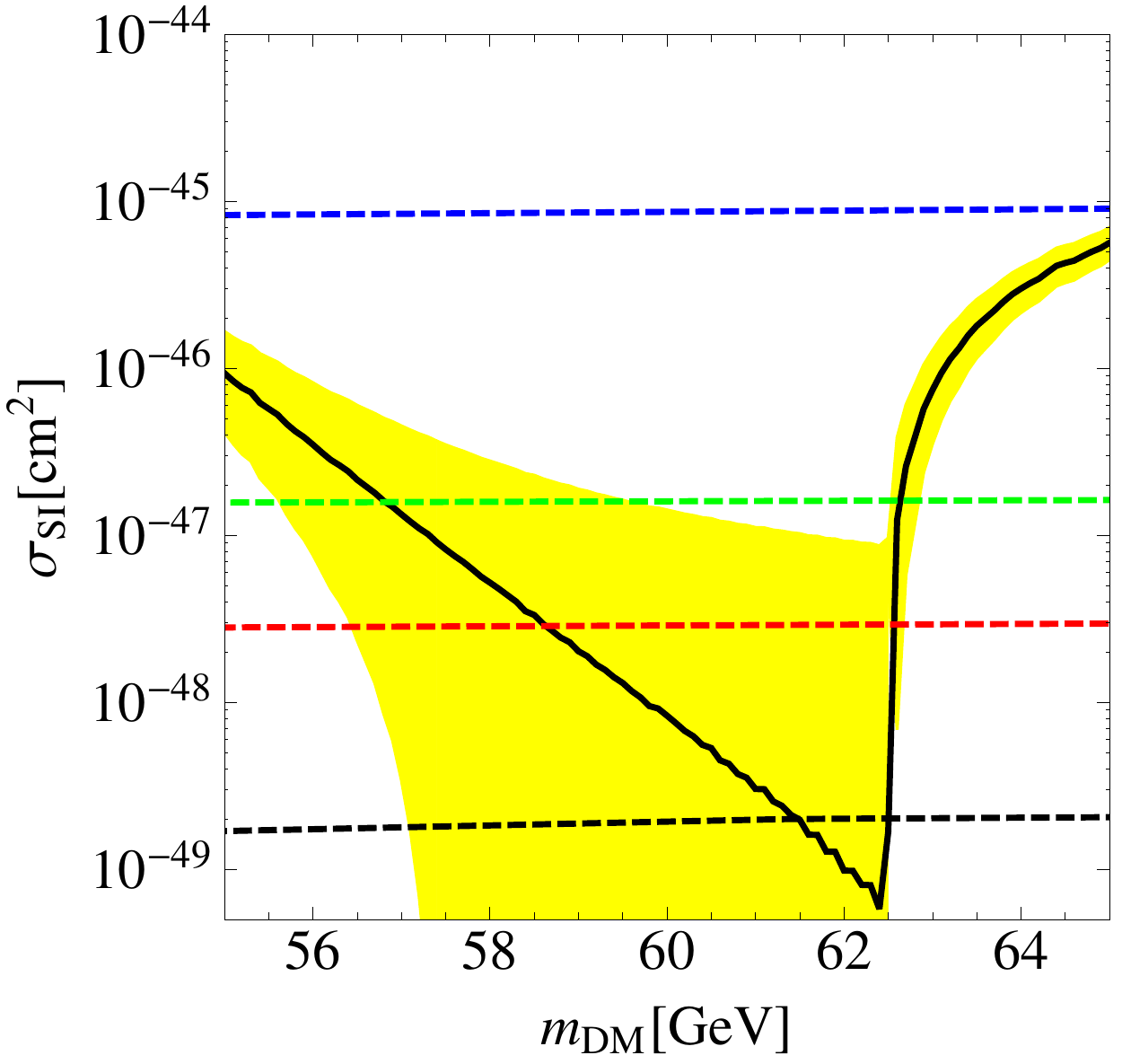} 
\caption{
The spin-independent cross section at tree level (black-solid line), and
 loop level (yellow shaded region). Here we vary $\lambda_2$ for $0 < \lambda_2 <
 1.45$. 
The meaning of the dashed lines are the same as in Fig.~\ref{fig:result1}.
Here we take $\Delta m_{H^{\pm}}=$
 50~GeV, $\Delta m_{S}=$50~GeV.
}
\label{fig:result2}
\end{figure}

\begin{figure}[tb]
  \includegraphics[width=0.32\hsize]{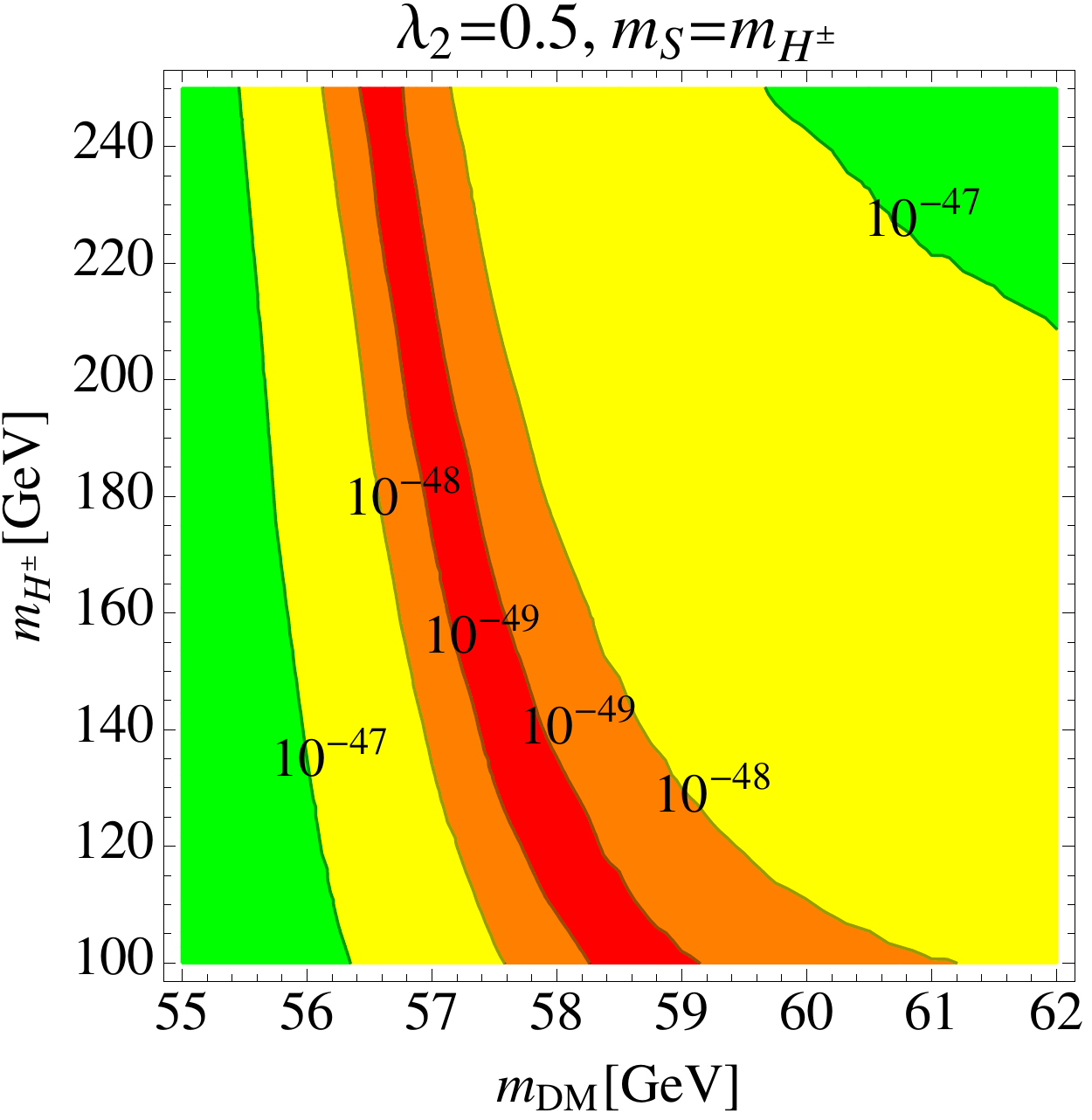} 
  \includegraphics[width=0.32\hsize]{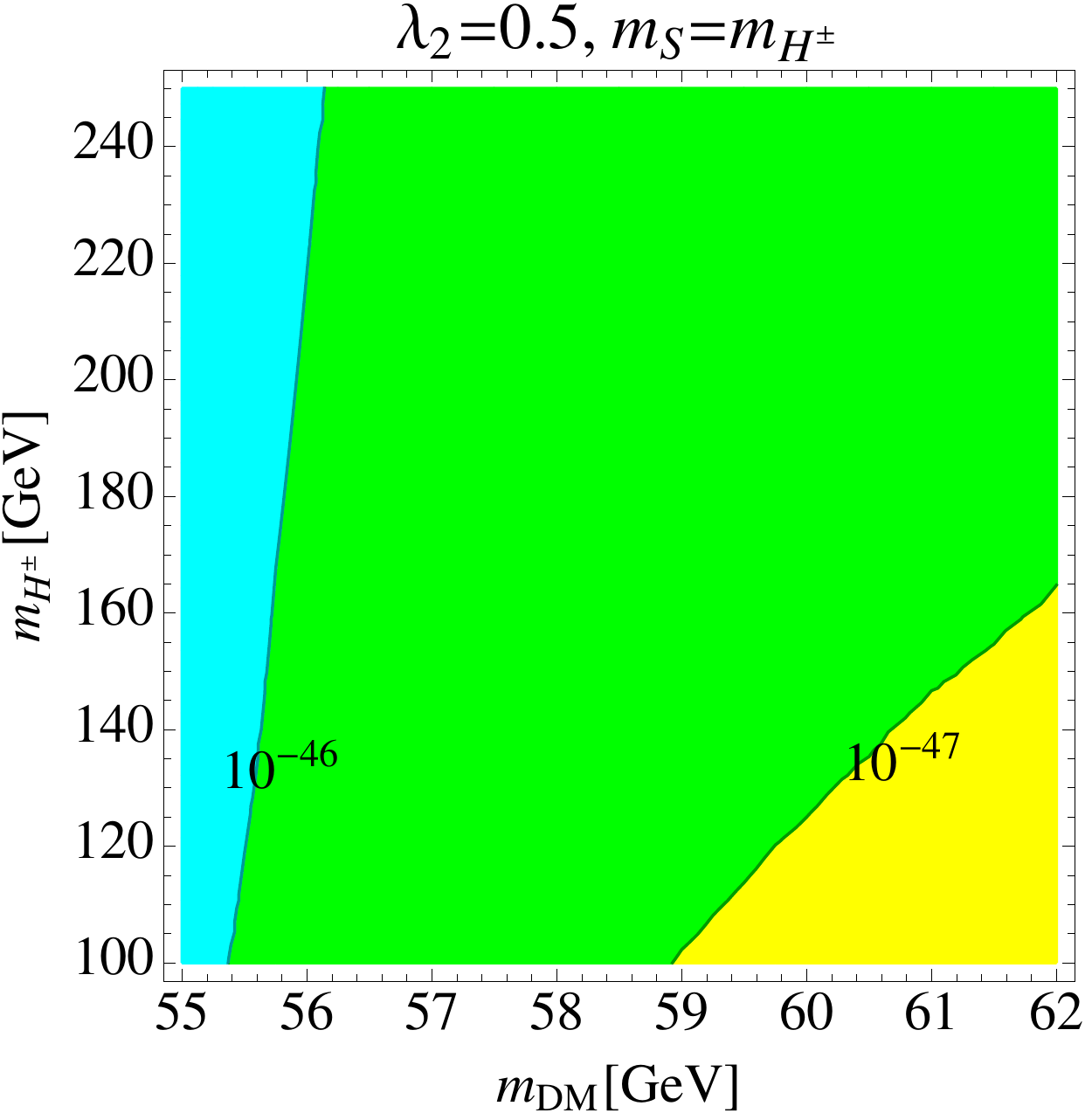} 
\caption{
The $\sigma_{\text{SI}}$ in ($m_{\text{DM}}, \lambda_2$)-plain.
The value on the panels is $\sigma_{\text{SI}}$ in cm$^2$ unit.
 In the left (right) panel, the sign of the $\lambda_A$ is
  positive (negative).
}
\label{fig:result3}
\end{figure}

We show the spin-independent cross section with and without the loop
correction in Fig.~\ref{fig:result1}. The difference in the three panels
is the choice of $\lambda_2$. We find that the loop correction is highly
depend on $\lambda_2$, and can be accidentally canceled out.
We vary the value of $\lambda_2$ for $0 < \lambda_2 < 1.45$
in Fig.~\ref{fig:result2}, because we do not know the value of
$\lambda_2$. The yellow region in the plot is the prediction of the
spin-independent cross section in this model. 
The dependence of the $m_{H\pm}$ and $m_{S}$ are shown in
Fig.~\ref{fig:result3}. We find that these two parameters also affect the prediction.

\section{summary}
We investigate the quantum corrections to the spin-independent cross
section in the inert-doublet model. The corrections are significant for
the light dark matter regime, 53~GeV$\lesssim m_{\text{DM}} \lesssim m_h/2$,
where the DM-Higgs couplings is highly suppressed. 
We find the loop
corrections significantly modify the prediction based on the tree
level analysis.  
At the tree level analysis, the spin-independent cross
section depends only on the dark matter mass after setting the DM-Higgs
coupling by the relic abundance. However, at the loop level, it depends
on the other model parameters as well. Especially, the effect from $\lambda_2$ and the sign
ambiguity of $\lambda_A$ are significant.

\begin{acknowledgments}
The work is supported by 
Grant-in-Aid for Scientific research from the Ministry of Education,
Science, Sports, and Culture (MEXT), Japan, No. 23104006 [TA]
and JSPS Research Fellowships for Young Scientists [RS].
\end{acknowledgments}

\bigskip 

\begin{thebibliography}{99} 
\bibitem{1501.04161} 
  T.~Abe and R.~Sato,
  JHEP {\bf 1503}, 109 (2015)
  [arXiv:1501.04161 [hep-ph]].


\bibitem{PHRVA.D18.2574} 
  N.~G.~Deshpande and E.~Ma,
  Phys.\ Rev.\ D {\bf 18}, 2574 (1978).


\bibitem{hep-ph/0603188} 
  R.~Barbieri, L.~J.~Hall and V.~S.~Rychkov,
  Phys.\ Rev.\ D {\bf 74}, 015007 (2006)
  [hep-ph/0603188].


\bibitem{LopezHonorez:2006gr} 
  L.~Lopez Honorez, E.~Nezri, J.~F.~Oliver and M.~H.~G.~Tytgat,
  JCAP {\bf 0702}, 028 (2007)
  [hep-ph/0612275].


\bibitem{JCAP.1406.030} 
  A.~Arhrib, Y.~L.~S.~Tsai, Q.~Yuan and T.~C.~Yuan,
  JCAP {\bf 1406}, 030 (2014)
  [arXiv:1310.0358 [hep-ph]].


\bibitem{Abe:2014gua} 
  T.~Abe, R.~Kitano and R.~Sato,
  arXiv:1411.1335 [hep-ph].


\bibitem{PRLTA.39.165} 
  B.~W.~Lee and S.~Weinberg,
  Phys.\ Rev.\ Lett.\  {\bf 39}, 165 (1977).


\bibitem{NUPHA.B310.693} 
  M.~Srednicki, R.~Watkins and K.~A.~Olive,
  Nucl.\ Phys.\ B {\bf 310}, 693 (1988).


\bibitem{NUPHA.B360.145} 
  P.~Gondolo and G.~Gelmini,
  Nucl.\ Phys.\ B {\bf 360}, 145 (1991).


\bibitem{Ade:2013zuv} 
  P.~A.~R.~Ade {\it et al.}  [Planck Collaboration],
  Astron.\ Astrophys.\  {\bf 571}, A16 (2014)
  [arXiv:1303.5076 [astro-ph.CO]].


\bibitem{PHRVA.D87.075025} 
  M.~Klasen, C.~E.~Yaguna and J.~D.~Ruiz-Alvarez,
  Phys.\ Rev.\ D {\bf 87}, 075025 (2013)
  [arXiv:1302.1657 [hep-ph]].


\bibitem{Carmona:2015haa} 
  A.~Carmona and M.~Chala,
  arXiv:1504.00332 [hep-ph].


\bibitem{1310.8214} 
  D.~S.~Akerib {\it et al.}  [LUX Collaboration],
  Phys.\ Rev.\ Lett.\  {\bf 112}, 091303 (2014)
  [arXiv:1310.8214 [astro-ph.CO]].


\bibitem{Aprile:2012zx} 
  E.~Aprile [XENON1T Collaboration],
  Springer Proc.\ Phys.\  {\bf 148}, 93 (2013)
  [arXiv:1206.6288 [astro-ph.IM]].


\bibitem{Feng:2014uja} 
  J.~L.~Feng, S.~Ritz, J.~J.~Beatty, J.~Buckley, D.~F.~Cowen, P.~Cushman, S.~Dodelson and C.~Galbiati {\it et al.},
  arXiv:1401.6085 [hep-ex].


\bibitem{Billard:2013qya} 
  J.~Billard, L.~Strigari and E.~Figueroa-Feliciano,
  Phys.\ Rev.\ D {\bf 89}, no. 2, 023524 (2014)
  [arXiv:1307.5458 [hep-ph]].


\end{thebibliography}

\end{document}